# Submicrometric Films of Surface-Attached Polymer Network with Temperature-Responsive Properties

M. Li,[1] B. Bresson,[1] F. Cousin,[2] C. Fretigny,[1] Y. Tran[1]*

[1] École Supérieure de Physique et de Chimie Industrielles de la Ville de Paris (ESPCI), ParisTech, PSL Research University, Sciences et Ingénierie de la Matière Molle, CNRS UMR 7615, 10 rue Vauquelin, F-75231 Paris cedex 05, France. Sorbonne-Universités, UPMC Univ Paris 06, SIMM, 10 rue Vauquelin, F-75231 Paris cedex 05, France

[2] Laboratoire Léon Brillouin, CEA-CNRS, Saclay 91191 Gif-sur-Yvette Cedex, France

* Address correspondence to Yvette.Tran@espci.fr

ABSTRACT.

Temperature-responsive properties of surface-attached poly(*N*-isopropylacrylamide) (PNIPAM) network films with well-controlled chemistry are investigated. The synthesis consists in crosslinking and grafting preformed ene-reactive polymer chains through thiol-ene click chemistry. The formation of surface-attached and crosslinked polymer films has the advantage of being well-controlled without any caution of no-oxygen atmosphere or addition of initiators. PNIPAM hydrogel films with same crosslinks density are synthesized on a wide range of thickness, from nanometers to micrometers. The swelling-collapse transition with temperature is studied by using ellipsometry, neutron reflectivity and atomic force microscopy as complementary surface probing techniques. Sharp and high amplitude temperature-induced phase transition is observed for all submicrometric PNIPAM hydrogel films. For temperature above LCST, surface-attached PNIPAM hydrogels collapse similarly but without complete expulsion of water. For temperature below LCST, the swelling of PNIPAM hydrogels depends on the film thickness. It is shown that the swelling is strongly affected by the surface attachment for ultrathin films below approximately 150 nm. For thicker films above 150 nm (to micrometers), surface-attached polymer networks with same crosslinks density swell equally. The density profile of the hydrogel films in the direction normal to the substrate is confronted with in-plane topography of the free surface. It results that the free interface width is much larger than the roughness of the hydrogel film, suggesting pendant chains at the free surface.

INTRODUCTION

Research interest in hydrophilic polymer coatings is due to their wide range of applications for materials, biology or medicine fields. For example, wetting, permeability as well as adhesive, adsorptive properties can be controlled and improved with appropriate polymers and especially stimuli-responsive polymers.[1-3] Polymer brushes and self-assembled layers (such as layer-by-layer assemblies) are so far polymer coatings most investigated in this area.[4-9] Despite wide and significant properties demonstrated in the past decades, these coatings have some inconvenience. The thickness of polymer brushes ruled by the polymer chain length is restricted to nanometer scale. The formation of layer-by-layer assemblies requires many cyclic steps (each layer is nanometer-thick) and so, high thickness could be the limitation. Moreover, if stimuli-responsive properties with high volume change are aimed, layer-by-layer assemblies are not the most suitable coatings.

Surface-attached hydrogel films are a promising alternative to polymer brushes and layer-by-layer assemblies as stable responsive polymer coatings.[10-11] A priori, the thickness of surface-attached polymer network would not be problematic as it could be widely ranged. Hydrogel films would respond to stimuli with high volume change. However, surface-attached hydrogel films with stimuli-responsive properties are still scarcely investigated. It is most likely due to a lack of common strategy for synthesis of hydrogel films with well-controlled chemistry. Actually, hydrogel films can be synthesized by polymerizing and crosslinking monomers by radical polymerization as for synthetic hydrogel materials. If molds (usually two substrates separated by spacers)[12] confining the reaction mixture (monomers, crosslinkers and initiators) are required, submicrometer-thick hydrogel films are unfeasible. In addition, the radical polymerization must be performed under controlled atmosphere to avoid oxygen. It results that the synthesis of hydrogel films at surface is very sensitive (to oxygen) due to a high surface to volume ratio. A way to overcome this difficulty is to preform functionalized polymers first and then simultaneously crosslink and graft chains to the surface. A few articles from Toomey et al.[13-16] and Kuckling et al.[17-20] have reported temperature-induced phase transition of poly(*N*-isopropylacrylamide) (PNIPAM) hydrogel films. Both groups cautiously synthesized surface-attached network films using photo-reactive PNIPAM. In detail, Kuckling used dimethylmaleimide groups as photo-crosslinkers and Toomey preferred benzophenone groups.

We propose another approach on the phase transition of surface-attached PNIPAM films and a new strategy of synthesis. In particular, we are interested in the swelling-collapse properties of PNIPAM hydrogel films on a large range of submicrometric thickness, which would help to determine the extent of the effect of surface attachment and confinement. This focus is important for some applications in which nanometric layers are involved as well as for other functional coatings with micrometric sizes. Another point concerns the synthesis of polymer network films. We propose a simple and versatile strategy based on thiol-ene click chemistry to crosslink polymer chains. Thiol-ene reaction[21] chosen here can be activated by thermal heating, which is supposed to guarantee a homogeneous distribution of crosslinks in the whole hydrogel film.

In this article, surface-attached PNIPAM network films are synthesized on a wide range of thickness from nanometer to micrometers. The temperature-induced phase transition is investigated using ellipsometry, neutron reflectivity and atomic force microscopy as complementary surface probing techniques. The effect of confinement and constraints (due to surface attachment) on the swelling-collapse phase transition of the polymer network films is investigated by two aspects: the one-dimensional swelling in the direction normal to the substrate and the in-plane observation of the free surface of the hydrogel. Ellipsometry allows the determination of the (average) swelling ratio. The range of thickness measured by ellipsometry is supposed to vary from nanometer to a few micrometers. From neutron reflectivity experiments, density profiles are deduced, providing the film thickness and the free interface width. The thickness range easily reachable is roughly from 1 nm (the limit is given by the range of wave vector) to 150 nm (the limit is given by the resolution of wave vector). The lateral (or in-plane) exploration of the free surface of the hydrogel films is achieved using AFM. The topography and the roughness of the free surface are confronted with the free interface width. It helps to estimate how diffuse (or smooth) is the free interface.

EXPERIMENTAL SECTION

**Synthesis of P(NIPAM-*co*-AA) copolymer.** Chemical products purchased in Sigma-Aldrich were used as received without any further purification (the purity ratio is more than 98% for all products). Poly(AA-*co*-NIPAM) was synthesized in Milli-Q pure water (18.2 MΩ.cm) by free radical polymerization initiated by $(NH_4)_2S_2O_8$ /$Na_2S_2O_5$ redox couple under nitrogen at room temperature. Once 53.5 mg $NH_4Cl$, 10.74 g NIPAM (0.095 mol) and 0.36 g AA (0.005 mol) are mixed in 100 ml water, the pH is adjusted to 3-4 with NaOH solution. Then the solution is deoxygenated with a bubbling of nitrogen for 1 h before the introduction of initiators (which were dissolved separately in 2 ml water and deoxygenated before). The polymer molecular weight is decided by the ratio of the $(NH_4)_2S_2O_8$ /$Na_2S_2O_5$ initiators. For example, a molecular weight of 250 kg/mol is obtained for the concentration of $(NH_4)_2S_2O_8$ and $Na_2S_2O_5$ in water equal to $10^{-3}$ mol/L. The reaction is allowed to proceed for 24 hours. The final solution is dialyzed in pure water and the polymer is recovered by freeze-drying.

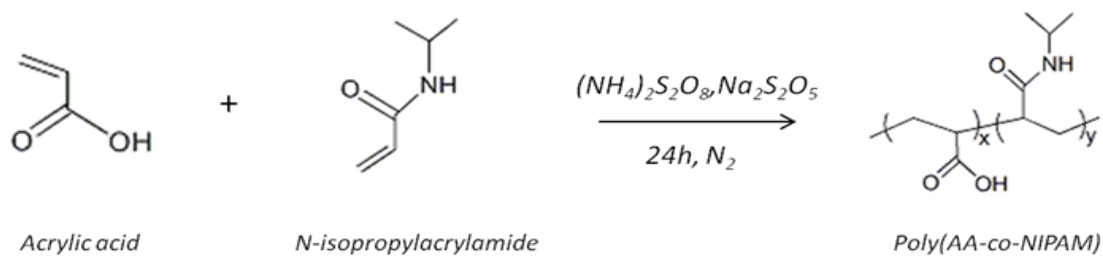

Acrylic acid     N-isopropylacrylamide     Poly(AA-co-NIPAM)

**Ene-functionalization of P(NIPAM-*co*-AA) copolymer.** The P(NIPAM-*co*-AA) copolymer is ene-functionalized by allylamine in Milli-Q water at room temperature in the presence of EDC/NHS couple. The molar ratio AA/allylamine/EDC/NHS is set equal to 1/2/2/2. After 5.4 g P(AA-*co*-NIPAM) (0.049 mol, if the ratio of AA in the copolymer is 5%) is dissolved in 150 ml water, EDC/NHS couple is added into the solution under stirring. The solution with a pH adjusted to 4.5 is left under stirring for 2h. The acidic conditions promote the protonation of EDC, facilitating the coupling reaction. Then, the aqueous solution of allylamine is added to the medium, associated with the adjustment of the pH to 10 with NaOH, for the formation of the targeted amide bond through the nucleophilic attack of allylamine onto the carboxylic acid. The reaction is allowed to proceed for 24 hours. The final solution is dialyzed firstly in 0.1 mol/L NaCl solution (to remove the unreacted EDC/NHS) for 4 days, and then in Milli-Q water for 4 days, solvent being changed twice a day. Finally the polymer is recovered by freeze-drying.

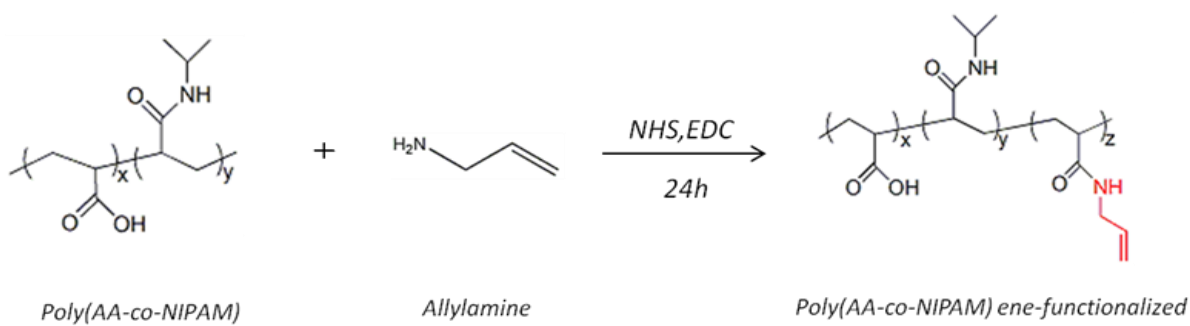

Poly(AA-co-NIPAM)  Allylamine  Poly(AA-co-NIPAM) ene-functionalized

**Thiol-modification of substrates.** Prior to any chemical modification of the silicon wafers, it is necessary to activate the silanol functions of the native layer of silica. The wafers are cleaned in a freshly prepared "piranha" solution (mixture of 70 vol% of sulfuric acid and 30 vol% of hydrogen peroxide - 35 vol% in $H_2O$ - heated at 150°C). The wafers are then extensively rinsed and sonicated in Milli-Q water during 1 min and finally dried with nitrogen flow. Silanization with 3-mercaptopropyltrimethoxysilane is carried out on the freshly cleaned silicon wafers for thiol-modification of the surface. After irradiated by UV-ozone for 15 minutes, the silicon wafers are quickly transferred into a sealed reactor filled with $N_2$. The solution of dry toluene with 3 vol% of mercaptopropyltrimethoxysilane (ABCR Gelest) is introduced into the reactor using cannula. The wafers are kept immersed in the solution under nitrogen for 3 hours. The samples are then rinsed and sonicated in toluene during 1 min and finally dried with nitrogen flow.

Before polymer coating, silicon substrates are thiol-modified with mercaptopropyltrimethoxysilane. The thickness of thiol layer measured by ellipsometry is 1 nm with a standard deviation lower than 10% of the average thickness. The characteristics of the thiol self-assembled monolayer are in good agreement with other fine studies (see for example Hu et al.[22]).

**Synthesis of hydrogel films**. Surface-attached hydrogel films are synthesized by simultaneously crosslinking and grafting reactive PNIPAM by thiol-ene reaction. Preformed ene-reactive (functionalized) PNIPAM is spin-coated on thiol-modified silicon wafers with dithiol crosslinkers. The mixture of butanol and methanol (V/V = 1/1) containing ene-reactive PNIPAM (at different concentrations and various molecular weight) and dithioerythritol crosslinkers is dropped onto thiol-modified solid substrates. The ratio of dithioerythritol to ene-functionalized copolymer units is 15 times (corresponding to a molar excess of bifunctional dithioerythritol of 30). The conditions of spin-coating are fixed with the final angular velocity of 3000 rpm and the spinning time of 30 seconds. The PNIPAM films are subsequently annealed at 120°C for 16 hours under vacuum to activate the thiol-ene reaction. The samples are rinsed and sonicated in water during 1 min and finally dried with nitrogen flow to remove free chains. The dry thickness (corresponding to the polymer amount) remains the same after many steps of immersion in solvent, demonstrating that the stability of polymer films is due to covalent bond between chains and with the substrate.

**Ellipsometry.** To measure the thickness, a spectroscopic ellipsometer Nanofilm EP3 (Accurion GmbH, Germany) was used with the wavelength ranging from 360 nm to 1000 nm. The resolution of $\psi$ and $\Delta$ ellipsometric angles is 0.001°. For the measurements in air, we used the model with two layers between two semi-infinite media which are the silicon substrate (refractive index equal to 3.87) and the ambient air (refractive index equal to 1). The first layer comprises silica and silane (refractive index equal to 1.46) which the thickness was determined before anchoring the hydrogel film. The second layer is the PNIPAM hydrogel film which thickness $h_a$ is measured with the refractive index of PNIPAM equal to 1.52. If we take in account that water content in PNIPAM film is 10% in air (water content lower than 10% was measured for humidity ratio between 20% and 60%), the refractive index of PNIPAM gel is 1.50. In fact, the water content is weak so that the thickness of PNIPAM gel film in air can be considered as the thickness of dry film. *In situ* measurements in water are performed using a liquid cell with thin glass walls fixed perpendicularly to the light path and the angle of incidence is fixed at 60°. The liquid cell is temperature-controllable with a regulation within ± 0.1°C. The PNIPAM hydrogel film is modeled as a single layer with the thickness $h_w$ and a constant refractive index between that of water (1.33) and of the polymer (1.52). Since the hydrogel film is covalently attached to the substrate, the polymer amount should keep the same when immersed in water. So we have: $n_w = (1.52 - 1.33) \times \phi_p^w + 1.33$ and $h_w \times \phi_p^w = h_a$ with $\phi_p^w$ the volume fraction of polymer in water. The swollen thickness $h_w$, the refractive index of the film in water $n_w$ and the dry thickness $h_a$ are deduced from the fitting of the experimental data. The two equations should provide the same value of $\phi_p^w$ to confirm that the fitting is reliable. Each data correspond to the mean measure of at least three samples, error bars providing the standard deviation from the mean value.

**Neutron reflectivity.** Neutron reflectivity measurements were carried out on the time-of-flight EROS reflectometer at the Laboratoire Léon Brillouin, CEA-Saclay (France). The wavelengths are in the range of 3 to 25 Å ($\Delta\lambda/\lambda \approx 0.1$) and the sample-to-detector distance is fixed at 2.0 m. The wave vector range from 0.005 Å$^{-1}$ to 0.1 Å$^{-1}$ is achieved with two grazing angles of 0.93° and 1.6° for the measures at air-silicon interface and 1.2° and 2° at silicon-D$_2$O interface. The beam size is adapted to the sample size and incident angle: 25 mm wide and 1 mm high. For the measures at silicon-water interface, the sample cell consists of the silicon single crystal and a temperature-regulated Teflon trough (filled with water) which are sandwiched in a stainless steel holder. PNIPAM hydrogels are protonated with a scattering length density equal to $1.0 \; 10^{-6} \; \text{Å}^{-2}$ (see discussion in Supporting Information) quite far from that of silicon ($2.1 \; 10^{-6} \; \text{Å}^{-2}$) and D$_2$O ($6.4 \; 10^{-6} \; \text{Å}^{-2}$), which provides good contrast for the measures in air and in water.

The reflectivity raw data is corrected for the direct beam and for the non-specular signal. It is then normalized using the position of the total reflectivity plateau. The logarithm of the analyzed reflectivity curves are plotted as a function of the wave vector $q = \frac{4\pi}{\lambda} \sin \theta$, where $\lambda$ is the wavelength and $\theta$ the grazing angle. The neutron reflectivity is sensitive to the scattering length density profile in the direction normal to the interface $Nb(z)$. The determination of $Nb(z)$ was achieved by modeling the hydrogel film as a set of three layers, each characterized by a fixed thickness $h_i$ and a fixed scattering length density $Nb_i$. Error functions of fixed width $\sigma_i$ allowed the connection of two adjacent layers to get a continuous profile. The scattering length density of the layer $i$ is written as $Nb_i(z) = Nb_i + \frac{Nb_{i+1} - Nb_i}{2} \left[ 1 + erf\left(\frac{z - h_i}{\sigma_i}\right) \right]$. The procedure consists of choosing a profile of scattering length density and finding the corresponding parameters for which the calculated reflectivity curve best fits the experimental reflectivity data. The model of three layers is for silica layer (with thickness between 1.5 and 2 nm in agreement with the values measured by ellipsometry) and thiol self-assembled monolayer (with thickness between 0.7 and 1 nm) as the two first layers. The third layer is the layer of interest as it corresponds to the PNIPAM hydrogel. The monomer volume fraction profile $\phi(z)$ which is the local monomer concentration at a distance $z$ from the interface, is deduced from $Nb(z)$ using $\phi(z) = \dfrac{Nb(z) - Nb_w}{Nb_p - Nb_w}$ where $Nb_w$ and $Nb_p$ are the scattering length density of $D_2O$ and PNIPAM (respectively equal to $6.4 \; 10^{-6}$ Å$^{-2}$ and $1.0 \; 10^{-6}$ Å$^{-2}$). The dry thickness of the polymer layer $h_a$ which corresponds to the integral of the monomer volume fraction profile $h_a = \int_0^{+\infty} \phi(z)\,dz$ is systematically compared to the value obtained by ellipsometry. The relation between the dry thickness $h_a$ and the swollen thickness $h_w$ as $h_w \times \phi_p^w = h_a$ ($\phi_p^w$ for the average volume fraction of polymer in water) is still used to judge the reliability of the data fitting.

**Atomic Force Microscopy.** All the images were acquired with a Bruker ICON microscope controlled by a Nanoscope V. The surface of hydrogel films in air can be easily probed with tapping mode. Cantilevers have a resonance frequency of about 300 kHz and a typical stiffness of 40 N/m. The working conditions were the following: free amplitude of about 20 nm, amplitude reduction equal to 0.95, scan frequency of 0.5 Hz and images acquired with 512 x 512 pixels. We checked that all the images were acquired in repulsive mode. To probe hydrogel films in water, we used the QNM mode (Quantitative Mechanical Measurements) where the cantilever height above the surface is periodic. This mode can be viewed as a succession of approach-retract curves at high frequency (1 kHz in our case), the amplitude being limited by the maximum set force. The QNM images in water were obtained using a cantilever of stiffness of 0.02 N/m, with a typical Force Set Point equal to 1 nN. The size of the images is 10 μm. The mean roughness is estimated from the half width of the height histogram of each image.

RESULTS AND DISCUSSION

**Preparation of PNIPAM network films**

Surface-attached polymer network films with well-controlled chemistry were prepared by using a simple and versatile approach based on click chemistry. Thiol-ene click reaction was chosen to simultaneously crosslink preformed polymer chains and graft them to surface. The preformed ene-reactive PNIPAM chains provide the temperature-responsive properties of the hydrogel films. Surface-attached hydrogel films are obtained by coating reactive polymer with dithiol crosslinkers on thiol-modified substrate (Figure 1). Thiol-ene reaction was selected for many reasons. It is advantageously performed without any added initiator and can be activated by thermal heating. The thermal activation rather than UV irradiation is supposed to guarantee homogeneous distribution of polymer crosslinks in the whole network film whatever the thickness. Many thiol molecules are commercially available for polymer crosslinking and surface functionalization. Dithiol (here dithioerythritol) was used as crosslinkers and mercaptosilane (here mercaptopropyltrimethoxysilane) as surface modifiers. Ene-reactive PNIPAM can be synthesized without difficulty and carefully characterized before coating. The synthesis of ene-reactive PNIPAM is carried out in two steps: (i) free radical polymerization of acrylic acid (AA) and *N*-isopropylacrylamide (NIPAM) and (ii) ene-functionalization of the copopolymer by amide formation using allylamine. The first stage is the free radical copolymerization of AA and NIPAM using ammonium persulfate/sodium metabisulfite redox couple as initiator. The molecular weight of polymer chains is ruled by the concentration of the reducing agent as shown by Bokias et al.[23] Free radical polymerization is the best option as facile polymerization technique because the control of the molecular weight distribution is not required as polymer chains are aimed to be crosslinked. The second step is the ene-functionalization of P(AA-*co*-NIPAM) by grafting allylamine in the presence of EDC/NHS couple, EDC being the dehydration agent and NHS the addition agent used to increase yields and decrease side reactions.[24] Both radical polymerization and amide formation are advantageously performed in water at ambient temperature as all the chemicals used are water-soluble. The ene-functionalized polymer can be easily purified by dialysis against water and recovered by freeze-drying to be characterized by Size Exclusion Chromatography and NMR spectroscopy. Three polymers were synthesized in this article. The molecular weight are equal to 66 kg/mol, 254 kg/mol, 405 kg/mol and 681 kg/mol with polydispersity index around 2. The ratio of ene-groups are equal to 2% (details are provided in Supporting Information).

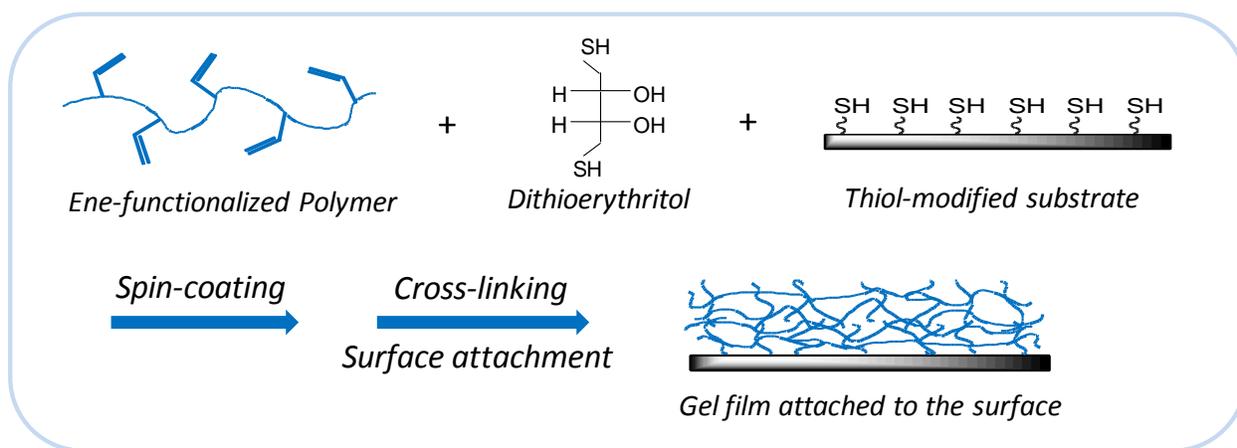

*Figure 1.* Schematic of hydrogel films synthesis: ene-reactive polymer is spin-coated on thiol-modified substrate with dithiol crosslinkers. Thiol-ene click reaction allows simultaneous polymer chains crosslinking and surface attachment.

Spin-coating was chosen as coating technique to synthesize polymer films with a wide range of thickness from nanometers to micrometers. The spin-coating technique has also the advantage to require a little polymer, for example, a polymer concentration of 1 wt% needs only 2 mg of polymer (one $cm^2$-surface can be covered by 0.2 ml of solution at the maximum). The solvent blend selected has to solubilize both ene-functionalized polymer and dithiol crosslinkers and also allow the spreading of polymer film on thiol-modified substrates by spin-coating. The mixture of butanol and methanol (V/V = 1/1) is the most suitable solvent blend.

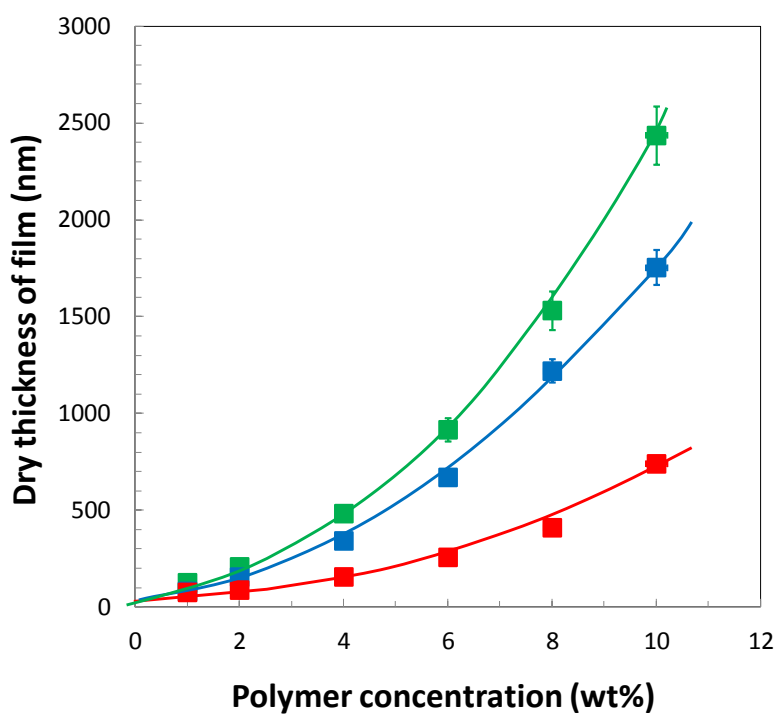

*Figure 2.* Dry thickness of PNIPAM hydrogel films as function of polymer concentration in the spin-coating solution for various molecular weights (66 kg/mol: red markers, 254 kg/mol: blue markers, 681 kg/mol: green markers, the solid lines being guides for the eye).

Figure 2 shows the variation of the thickness of surface-attached PNIPAM network film with the change of polymer concentration and molecular weight. This calibration curve gives evidence that the thickness can widely range from nanometers to micrometers using spin-coating technique. As expected, the film thickness increases with the viscosity of the polymer solution, and so with polymer concentration or molecular weight. For molecular weight of 66 kg/mol, submicrometer thickness can be obtained with a good precision (below 5% deviation) by varying the concentration below 10 wt%. If higher molecular weight is used, the films have a bigger deviation of the thickness (up to 10%) but the advantage is the large range with the formation of micrometric films.

**Temperature-induced phase transition**

The temperature-induced phase transition of surface-attached PNIPAM hydrogel films was investigated by measuring the thickness of the hydrogel films in water at various temperatures (Figure 3). The swelling ratio is defined as the thickness of hydrogel film in water (swollen thickness) to that in air (dry thickness). The data are obtained with hydrogel films which are supposed to have same crosslinks density. This assumption seems reliable as the synthesis of hydrogel films is performed with the same reactive PNIPAM (2% ene-functionalized) and same excess of dithioerythritol cross-linkers (30 times the ene-reactive groups). If same experimental conditions were used except the thickness, hydrogel films can be supposed to be the same chemically. The crosslinking is likely uniform without any gradient in the direction normal to the surface as ene-functionalized polymers and dithiol-crosslinkers are homogeneously mixed before spin-coating. It is not possible to determine the crosslinks density, and specially the crosslinking gradient, by spectroscopic surface probing techniques such as X-ray Photoelectron Spectroscopy (XPS) or Infrared Spectroscopy in Attenuated Total Reflection (FTIR-ATR). As XPS technique can only probe nanometric layers (the standard penetration depth is about 5 nm), it cannot provide the chemical composition of the whole hydrogel films. FTIR-ATR method with micrometric penetration depth could be more appropriate. Unfortunately, the crosslinks ratio is too weak to be quantified by ATR, the maximum being 2% (which corresponds to the ratio of ene-reactive groups determined by $^1$H NMR). Anyway, the absorption peak which is characteristics of thiol-ene reaction (S-C bonds) is positioned about 700 cm$^{-1}$, i.e. in the wave number range of high absorbance of the silicon waveguide (below 1600 cm$^{-1}$). In fact, the measure of the thickness in the swollen state is a way to control the synthesis of hydrogel films since it is related to the crosslinks density. The swelling ratio is higher for weaker crosslinks density.

Unlike macroscopic polymer networks, surface-attached hydrogel films are expected to swell in one direction normal to the substrate (the dimension of the lateral surface is infinite compared to the perpendicular direction). As shown by Toomey et al.[13] the linear swelling of surface-attached hydrogels can be simply described by Flory-Rehner theory[25] for the swelling of hydrogels extended to one dimension. The degree of swelling in surface-attached network is a power law of the degree of swelling in the free network with the exponent 5/9, and not 1/3. They determined the swelling ratio (or linear degree of swelling) as function of the proportion of the photo-crosslinkable units from 0.5% to 14.3% for 100 nm-thick films. We found a swelling ratio of 4 found with 2% ene-functionalized polymer. The same value of 4 was found by Toomey et al. with 2% of photo-crosslinkable units. It means that thiol-ene chemistry and photo-crosslinking strategy provide same crosslinks density of surface-attached PNIPAM hydrogel films, which is 2% at the maximum. The swelling ratio of 4 obtained at ambient temperature (below the LCST of PNIPAM) is independent of the thickness in the whole range from 100 nm to 1 μm, indicating that the crosslinks density is probably identical for all these submicrometer PNIPAM films.

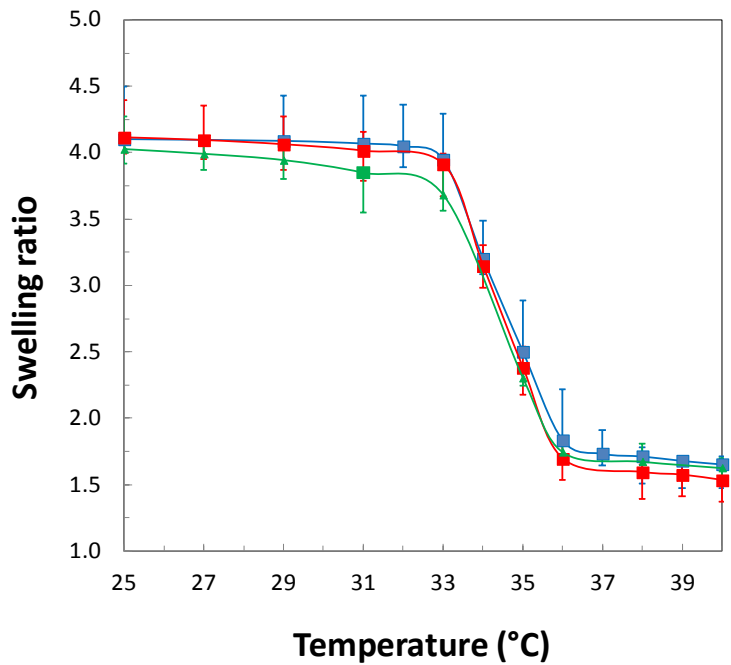

*Figure 3.* Swelling ratio of PNIPAM hydrogel films as function of temperature for various dry thickness of the films (using PNIPAM chains with the molecular weight equal to 254 kg/mol). Blue data: 150 nm, red data: 250 nm, green data: 420 nm, solid lines are guides for the eye.

For high temperatures above the LCST, the value of swelling ratio of 1.5 is found. Water is not entirely expulsed from the collapsed PNIPAM hydrogel which roughly 30% of water. It is in good agreement with the results from Kuckling[17-20] and Toomey[14]. They found that collapsed PNIPAM hydrogels are not free of water no matter the crosslinks density. It was not only observed for collapsed hydrogels but also for collapsed PNIPAM brushes by Yim et al.[26-29] The comparison of the swelling ratios of the PNIPAM hydrogel in the swollen and collapsed states shows that the phase transition is quite high amplitude (almost 3), which is interesting for applications using thermo-responsive properties. Another point which could be attractive for applications is the sharp transition within a temperature range of 3°C (from 33°C to 36°C). The same sharp transition with high amplitude is observed for surface-attached PNIPAM films in the whole range of thickness from 100 nm to 1 μm.

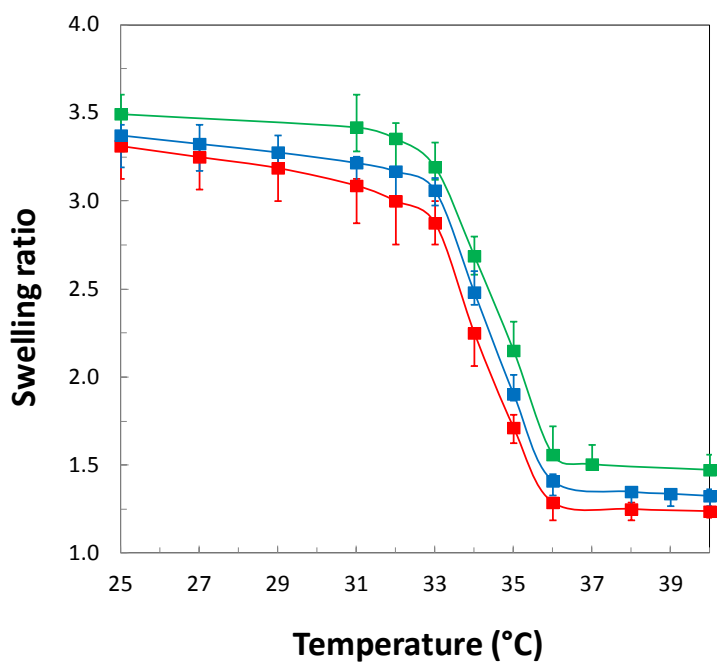

*Figure 4.* Swelling ratio of PNIPAM hydrogel film as function of temperature. The data are shown for polymers with various molecular weight used for the synthesis of the hydrogel films. Blue data: 66 kg/mol (the dry thickness of the film is 93 nm), red data: 405 kg/mol (80 nm), green data: 681 kg/mol (83 nm), solid lines are guides for the eye.

The sharp transition was also found for PNIPAM hydrogels synthesized from different molecular weight, as shown in Figure 4. Surface-attached hydrogels of similar thickness around 90 nm obtained with molecular weight varying from 66 kg/mol to 681 kg/mol were investigated. The same temperature-induced phase transition was observed. If the sharp transition is identical to that displayed in Figure 3, the amplitude of the transition is slightly different. The swelling ratio for collapsed hydrogels at high temperature above the LCST is around 1.5 (±0.2), which is similar to the value found for thicker films as shown in Figure 3. The difference is the swelling ratio at ambient temperature which is lower than 4. If PNIPAM hydrogel films swell less while they are supposed to have the same crosslinks density, it could be explained by the effect of the surface attachment, as demonstrated in the next part.

**Density profile determined by neutron reflectivity**

Ultra-thin films below 150 nm were investigated by neutron reflectivity. The thickness range easily reachable is roughly from 1 nm to 150 nm. The inferior limit is fixed by the range of wave vector and the superior limit by the resolution of wave vector. Also, we chose neutron reflectivity technique to characterize PNIPAM hydrogel films which thickness in air (respectively in water) varies from 6.7 nm to 36.4 nm (respectively up to 110 nm). Neutron reflectivity allows the determination of the density profile, i.e. the thickness and the width of the free surface of the hydrogel film.

Figure 5 shows the experimental reflectivity curves and their best fitting as well as the corresponding density profiles of PNIPAM hydrogel films at various temperatures. Temperatures are fixed at 25°C (below the LCST), 34°C (within the transition range) and 40°C (above the LCST). As temperature increases, the reflectivity curve displays more obvious Kiessig fringes, suggesting that corresponding density profile is sharper. Experimental data can be satisfactorily fitted with a model of one layer for PNIPAM film (a model of two or more layers do not allow the improvement of the fitting). It should be noticed that it changes from the study of polymer brushes by neutron reflectivity where the data fitting requires model with several layers for the smooth density profile of polymer brushes.[26-32] In Figure 5 with the density profiles, it can be observed that the thickness of PNIPAM hydrogel decreases with temperature, which is consistent with the temperature-induced phase transition. The average volume fraction is close to 0.3, which corresponds to a swelling ratio of 3 for the swollen hydrogel at 25°C. The volume fraction is a little higher than 0.6, indicating a water content of about 30% in the collapsed hydrogel at 40°C. The density profiles are comparable to those obtained by Toomey et al. on 25 nm-thick PNIPAM hydrogel.[14] Moreover, it should be noticed that the free interface width of of the hydrogel decreases with temperature. It is observed for surface-attached PNIPAM hydrogel films of any thickness. Reflectivity curves and density profiles of PNIPAM films with thickness in air equal to 7 nm and 12 nm are displayed in Supporting Information. All the parameters extracted from the fitting of neutron reflectivity curves which are characteristics of PNIPAM hydrogel layers (thickness in air and in water, volume fraction, interface width…) are additionally compiled in Supporting Information.

The same tendency described above is noted for the three hydrogel samples no matter the thickness: decrease of the thickness in water with temperature, swelling ratio between 2.5 and 3 for the swollen hydrogel at 25°C, water content of about 30% in the collapsed hydrogel at 40°C. For experiments in air, the interface width $\sigma$ increases with the hydrogel thickness $h$ but if it is normalized by the thickness, the relative interface width is the same for any thickness and is quite weak ($\sigma/h < 10\%$). The relative interface width is higher in water than in air. It varies with temperature and also with film thickness as it is weaker for thicker films. At 25°C, $\sigma/h$ decreases from 0.36 (for 6.7 nm-thick film measured in air) to 0.17 (for 36.4 nm-thick film). At 40°C, $\sigma/h$ decreases from 0.25 (for 6.7 nm-thick film) to 0.06 (for 36.4 nm-thick film). Surface-attached PNIPAM hydrogels look like more diffuse at the free surface when they are in the swollen state below the LCST than in the collapsed state above the LCST. This point is discussed in detail in the next part.

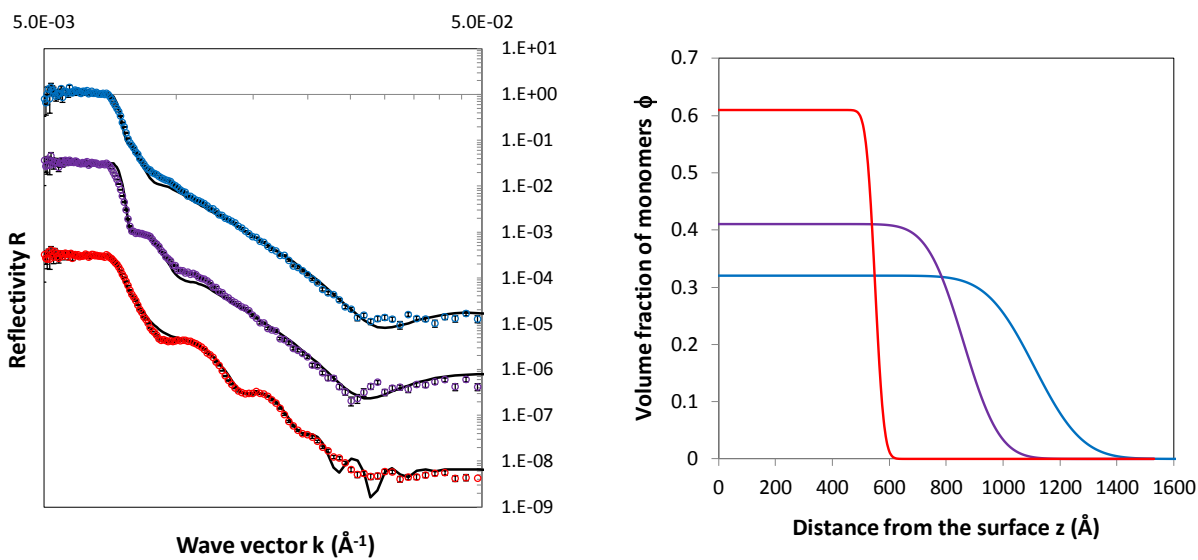

**Figure 5.** (Left) Neutron reflectivity curves of PNIPAM hydrogel film (with dry thickness equal to 36.4 nm) in water at three temperatures 25°C, 34°C and 40°C. Also are shown the best fitting of the experimental data. Circles are for data and lines for fitting. (Right) Volume fraction profiles corresponding to the best fitting. The same color code as the left figure with reflectivity curves is used: blue for 25°C, purple for 34°C and red for 40°C.

**Hydrogel films more diffuse than rough**

As ellipsometry, specular neutron reflectivity provides information on the slice of the film or in the direction normal to the substrate considering the composition of the x-y plane homogeneous. Actually, neutron reflectivity cannot distinguish a diffuse film from a rough film if the size of the in-plane irregularities is much smaller than the coherent length of neutrons. For thermal neutrons (with a typical wavelength of a few Ångström), the coherent length is a few microns. It is then helpful to compare the roughness of the surface of the gel film determined by AFM and the free interface width deduced from neutron reflectivity.

The height images of surface-attached PNIPAM gel films are displayed on Figure 6. The surface of the hydrogel film is probed in air with tapping mode. The QNM images were obtained in water using a cantilever of stiffness of 0.02 N/m with a typical force set-point equal to 1 nN. The size of the images is 10 micrometers. The mean roughness is estimated from the half width of the height histogram of each image. Precision on AFM imaging conditions is provided in Supporting Information. In the dry state (in air), the roughness of the film is subnanometric, which is small in comparison with the thickness of the film equal to 250 nm. When immersed in water at 25°C (< LCST), the roughness slightly increases to 5 nm with the vertical swelling of the hydrogel. At 40°C (> LCST), it is about 3 nm as the PNIPAM hydrogel collapses. Although the roughness (slightly) changes, it remains much lower than the film thickness so that PNIPAM hydrogel films can be considered as flat whatever the conditions in air and in water. The morphologies shown here are quite different from the instabilities observed by Toomey et al. with photo-crosslinked PNIPAM coatings.[16] They analyzed blisterlike or foldlike (also called *sulci*) instabilities resulting from the free surface buckling when surface-attached hydrogel films synthesized in dry state were exposed to solvent. Actually, in-plane swelling is frustrated in constrained hydrogel films. As the swelling occurs perpendicularly to the constraining surface but not laterally, in-plane compressive stress is generated. Under sufficient stress, the free surface becomes unstable and deforms out of plane. They demonstrated that the characteristics of the instabilities such as the wavelength, the width and the amplitude increase with the film thickness. For example, they found amplitude of 10 nm for 250 nm-thick PNIPAM film exposed to water, which is a little higher than our roughness value (5 nm). Indeed, the morphology shown in Figure 6 is similar to that obtained by Toomey et al. for thinner films. The difference is probably due to the synthesis of surface-attached PNIPAM hydrogel films. In our case, the crosslinking of polymer chains and their attachment to the surface are thermally activated instead of photo-activation. The annealing occurs at high temperature (120°C) so that the crosslinking is homogenously distributed in the whole film. In contrast, the photo-crosslinkings is performed at ambient temperature in the glassy PNIPAM layer, leading to disparity in crosslinking (or modulus) of the polymer network and promoting surface instabilities.

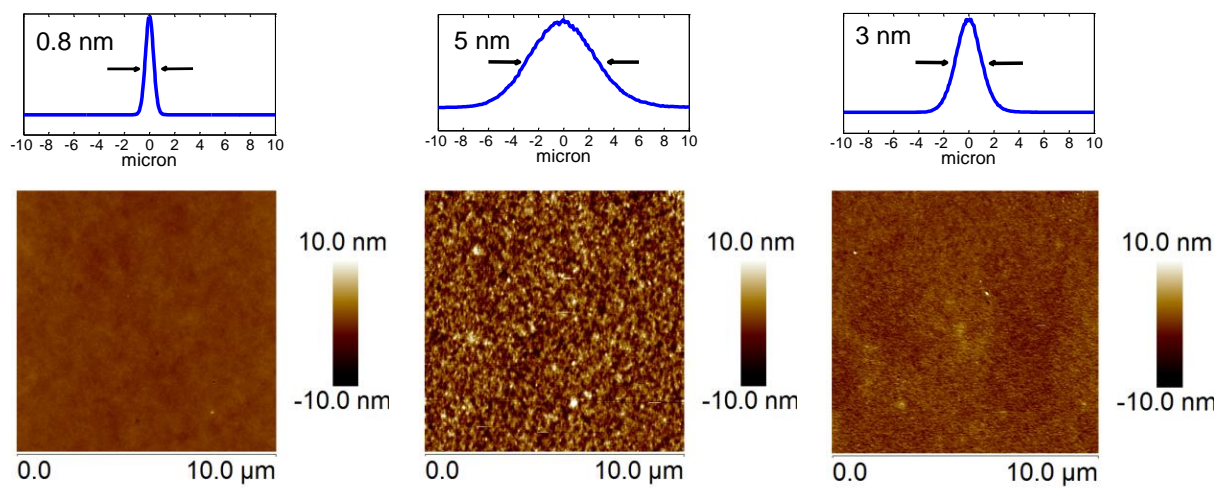

**Figure 6.** Height AFM images of PNIPAM hydrogel films in air (left), in water at 25°C (middle) and in water at 40°C (right). Images are obtained in air with tapping mode and in water by using QNM mode with a cantilever of stiffness 0.02 N/m. The thickness of PNIPAM hydrogel film is 250 nm in air, it is equal to 960 nm in water at 25°C and 360 nm at 40°C. The corresponding height histogram is displayed on the top. Roughness is determined from the half width of the peak.

| Samples | $h$ (nm) | $\sigma$ (nm) | $\rho$ (nm) | $\sigma/h$ or $\rho/h$ |
|---|---|---|---|---|
| **SN36-air** | 36.4 | 3.0 | | 8% |
| **SN36-25°C** | 110.7 | 18.8 | | 17% |
| **SN36-40°C** | 55.5 | 3.5 | | 7% |
| **SN36-air** | 36.0 | | 0.8 | 1.6% |
| **SN36-25°C** | 110.0 | | 1.1 to 2.0 | 1.8% |
| **SN250-air** | 250.0 | | 0.8 | 0.3% |
| **SN250-25°C** | 960.0 | | 5 | 0.5% |
| **SN250-40°C** | 360.0 | | 3 | 0.8% |

**Table 1.** Main characteristics extracted from AFM and neutron reflectivity experiments. The samples are named as SNx-y where x stands for thickness in air and y for experimental conditions. Are reported the thickness $h$ of the film, the free interface width $\sigma$ deduced from neutron experiments and the mean roughness $\rho$ determined by AFM. The relative interface width $\sigma/h$ and relative roughness $\rho/h$ are also calculated.

The main characteristics extracted from AFM and neutron reflectivity experiments are summarized in Table 1. The free interface width $\sigma$ deduced from neutron experiments and the mean roughness $\rho$ determined by AFM, are compared with normalization to the film thickness $h$. The relative roughness $\rho/h$, is much lower than the relative interface width $\sigma/h$, in any condition and in particular in water a 25°C. Note that $\rho/h$ values shown here for the 36 nm-thick film are the smallest in comparison with those reached with thinner films (6 nm- and 12 nm-thick films). The gap between $\rho/h$ and $\sigma/h$ is the proof that the smoothness of the free surface is not only due to the (in-plane irregularities) roughness but also to the diffusion of the interface. The free surface is the most diffuse when the surface-attached hydrogel is swollen. The diffuse surface might be explained by the presence of pendant chains at the outside edge of the film. The peripheral chains are more likely to penetrate into the aqueous environment (rather than into the network, which is entropically unfavorable) giving rise to a diffuse interface.

**Effect of the surface attachment**

Ellipsometry and neutron reflectivity allows the determination of the (average) swelling ratio of the hydrogel films in the direction normal to the substrate. The thickness range easily reachable by neutron reflectivity experiments is roughly below 150 nm as the limit is given by the resolution of wave vector. We used ellipsometry to characterize the swelling of hydrogel films which thickness in water ranges roughly from 300 nm to 1500 nm. In this range, the number of oscillations, for $\psi$ and $\Delta$ ellipsometric angles *versus* wavelength, is high enough to validate the relevance of experimental data fitting (for ultra-thin hydrogel layers, the oscillations are not obvious and especially in water with very smooth free interface).

In Figure 7, the swelling ratio of surface-attached PNIPAM hydrogel films is plotted as function of the dry thickness (in air). Data shown for three temperatures (25°C, 34°C and 40°C) were extracted from ellipsometry and neutron reflectivity experiments which combination allows the plot on a large range of thickness from nanometer to micrometer. At 40°C for collapsed gels, the swelling ratio is 1.5 on the whole range of thickness, which was already discussed. Collapsed PNIPAM hydrogel films are not free of water and contain roughly one third of water. At 25°C, the swelling properties of hydrogel films show clearly two regimes with a changeover around 150 nm. For thin films below 150 nm, the swelling ratio increases (from 2.5 to 4) with the film thickness and it is equal to 4 above 150 nm. The same is found at 34°C with lower swelling ratio (from 2 to 3). If the distribution of crosslinks is supposedly homogeneous in the whole hydrogel film for any thickness, it gives evidence that the surface attachment has a strong effect on the swelling of films. Nanometric hydrogel films swell on average less than micrometric layers as a result of the constraint due to the surface attachment. Actually, all concentration profiles measured by neutron reflectivity show a very dense layer next to the surface which is additional to the swollen hydrogel layer (data not shown). However, it is tricky to determine the characteristics of this dense layer with high precision as it is blended with thiol monolayer (of equivalent scattering length density). The additional layer which is required for the data fitting of hydrogel films in water (but no needed for samples in air) is roughly one nanometer-thick.

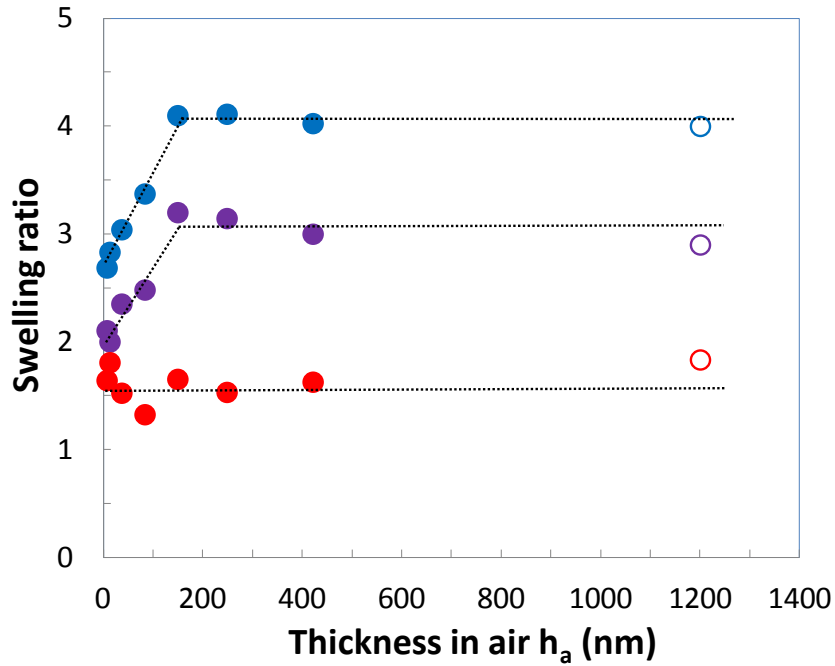

**Figure 7.** Swelling ratio of PNIPAM hydrogel films as a function of the film thickness in air, at three different temperatures: 25°C (blue data), 34°C (purple data) and 40°C (red data). Data with $h_a$ < 100 nm are from neutron reflectivity experiments, those between 100 nm and 1 μm are provided by ellipsometry and the data above 1 μm are from microfluidics measurements.

The data reported for 1.2 micrometer-thick films were obtained by using microfluidics technology. Briefly, the thickness of the hydrogel film was deduced from the measurements of the hydrodynamic resistance through microchannels in which surface-attached hydrogels were embedded. Although microfluidics experiments are not the purpose of this article, these measurements are intentionally mentioned to demonstrate that the regime of unvarying swelling is also valid for micrometric films.

CONCLUSIONS

Submicrometric films of surface-attached PNIPAM network films with well-controlled chemistry were synthesized with a new and simple strategy based on thiol-ene click chemistry. The strategy consisted in preforming ene-functionalized polymer chains first and then simultaneously crosslinking and grafting the chains to the surface by thiol-ene reaction. The thermal activation of thiol-ene reaction is supposed to ensure that crosslinks are homogeneously distributed in the whole hydrogel film whatever the thickness. Temperature-induced phase transition of PNIPAM hydrogel films was investigated by determining the one-dimensional swelling in the direction normal to the substrate and in-plane topography of the free surface. The maximum of swelling is obtained for temperature below 32°C and the maximum of collapse above 35°C. It is shown that collapsed PNIPAM hydrogels are not free of water (it contains roughly one third of water). Two regimes of swelling were clearly shown. In the first regime, the average swelling of PNIPAM hydrogels is strongly affected by the surface attachment for ultrathin films below approximately 150 nm with less swelling on average for thinner films. Thick films above 150 nm swell equally. The comparison of the topography of the free surface and the density profiles demonstrates that the hydrogel layer is more diffuse than rough suggesting existence of pendant chains at the free surface. This study is the starting point for applications exploiting stimuli-responsive properties of hydrogel thin films such as in microfluidic devices.

SUPPORTING INFORMATION AVAILABLE. 1/ Characterization of ene-functionalized polymers by $^1$H NMR spectroscopy. 2/ Characterization of PNIPAM hydrogel films by neutron reflectivity: additional experimental and fitted curves as well as corresponding density profiles. 3/ Characterization by AFM: precision on AFM imaging conditions.

ACKNOWLEDGMENT. We sincerely thank Ralph Colby for kind and fruitful discussions. The research was financially supported by the French National Research Agency (ANR) and the China Scholarship Council (CSC).

REFERENCES.

TABLE OF CONTENTS

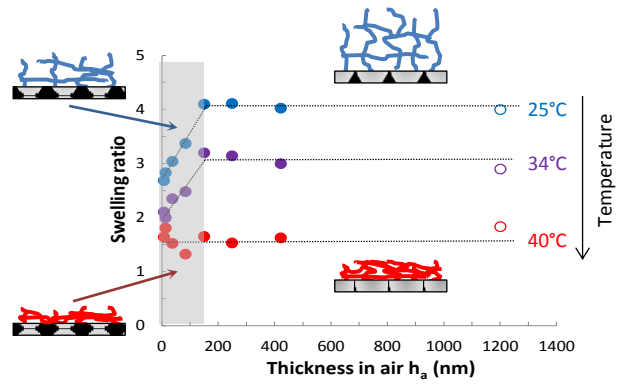